\documentclass[aps,pre,twocolumn,amsmath,amssymb,a4paper]{revtex4-1}

\usepackage{graphicx}
\usepackage{mathtools}
\usepackage{color}
\usepackage{tabularx}
\usepackage{float}

\usepackage{siunitx}
\usepackage{gensymb}

\definecolor{cream}{RGB}{222,217,201}

\begin{document}

\title{Coarse-Grained Nucleic Acid - Protein Model for Hybrid Nanotechnology}
\date{\today}
\author{Jonah Procyk$^{1}$}
\author{Erik Poppleton$^{1}$}
\author{Petr \v{S}ulc$^{1}$}
\email[Corresponding author: ]{psulc@asu.edu}
\affiliation{
$^1$School of Molecular Sciences and Center for Molecular Design and Biomimetics, The Biodesign Institute, Arizona State University, 1001 South McAllister Avenue, Tempe, Arizona 85281, USA \\
}

\begin{abstract}
 The emerging field of hybrid DNA - protein nanotechnology brings with it the potential for many novel materials which combine the addressability of DNA nanotechnology with versatility of protein interactions. However, the design and computational study of these hybrid structures is difficult due to the system sizes involved. To aid in the design and in silico analysis process, we introduce here a coarse-grained DNA/RNA-protein model that extends the oxDNA/oxRNA models of DNA/RNA with a coarse-grained model of proteins based on an anisotropic network model representation. Fully equipped with analysis scripts and visualization, our model aims to facilitate hybrid nanomaterial design towards eventual experimental realization, as well as enabling study of biological complexes. We further demonstrate its usage by simulating DNA-protein nanocage, DNA wrapped around histones, and a nascent RNA in polymerase.
\end{abstract}
\maketitle





\section*{Introduction}
Molecular nanotechnology designs biomolecular interactions to assemble nanoscale devices and structures. DNA nanotechnology, in particular, has attracted lots of attention and experienced rapid growth over the past three decades. While originally envisioned as a method of developing a DNA lattice for crystallizing proteins for structure determination \cite{Seeman1982}, DNA nanotechnology is seeing promising applications in  e.g.~biomaterial assembly \cite{liu2016diamond}, biocatalysis \cite{Geng2014}, therapeutics \cite{li2018dna}, and diagnostics \cite{Zhang2014}. The programmability of DNA allows for the rapid design and experimental realization of complex shapes, yielding an unprecedented level of control and functionality at the nanoscale. As DNA nanotechology has developed, so have parallel technologies with other familiar biomolecules such as RNA \cite{guo2010emerging}, and, to some extent, proteins \cite{Ulijn2018,king2014accurate}. While DNA nanostructures and devices have been unequivocally successful in realizing more complex and larger constructs, they are inherently limited in function by their available chemistry, with a possible solution using functionalized DNA nanostructures \cite{madsen2019chemistries}.
Of particular interest is hybrid DNA-protein nanotechnology, which can combine the already well developed design strategies of DNA nanotechnology and cross-linking them with functional proteins. The combination of the two molecules in nanotechnology will open new applications, such as diganostics, therapeutics, molecular "factories" and new biomimetic materials \cite{stephanopoulos2020hybrid}. Examples of successfully realized hybrid nanostructures include DNA-protein cages \cite{Xu2019}, a DNA nanorobot with nucleolin aptamer for cancer therapy \cite{Li2018} and peptide-directed assembly of large nanostructures \cite{jin2019peptide}.

%

At the same time, computational tools for the study and design of DNA and RNA nanostructures have become increasingly relevant as size and complexity of nanostructures grow. Design tools such as Adenita \cite{DeLlano2020} MagicDNA \cite{magicDNA}, CadNano \cite{Douglas2009}, and Tiamat \cite{williams2008tiamat} are essential for the structural design of DNA origamis.
New coarse-grained models have been introduced to study DNA nanostructures, as the sizes (thousands or more) as well as rare events (formation or breaking of large sections of base pairs) involved in study of these systems make atomistic-resolution modeling more difficult. Among the available tools, the oxDNA and oxRNAs models \cite{ouldridge2011structural,snodin2015introducing,vsulc2014nucleotide,sulc2012sequence} have been among the most popular over the past few years and have been used by dozens of research groups in over one hundred articles to study various aspects of DNA and RNA nanosystems as well as biophysical properties of DNA and RNA \cite{Sharma2017, Engel2018, Suma2019, Hong2018, Doye2013,matthies2019triangulated}. Each nucleotide is represented as a rigid body in the simulation, with interactions between different sites parametrized to reproduce mechanical, structural and thermodynamic properties of single-stranded and double-stranded DNA and RNA respectively.

However, the oxDNA/oxRNA models only allow for representation of nucleic acids alone, limiting their scope of usability. While there have been examples of coarse-grained protein-DNA modeling specifically applied to modeling of nucleosomes \cite{zhang2016exploring,honorato2019martini}, we currently do not have an efficient tool at the level of oxDNA coarse-graining that would allow for efficient study of arbitrary protein-DNA complexes. 

Here, we introduce such a coarse-grained model that uses an Anisotropic Network Model (ANM) to represent proteins alongside the oxDNA or oxRNA model. The ANM is a form of elastic network model used to probe the dynamics of biomolecules fluctuating around their native state. Originally formulated by Atilgan et. al. \cite{Atilgan2001}, the ANM has become fundamental tool in probing protein dynamics, often closely matching residue-residue fluctuations and normal modes of fully atomistic simulations \cite{Mishra2018,Gur2013,Yang2008}. Here we use the ANM to approximately capture native state protein dynamics. The ANM representation of proteins interact with just an excluded volume interaction with the oxDNA / oxRNA representation, but specific attractive or repulsive interactions can be added as well. We provide parametrization of common linkers that are used to conjugate proteins to DNA in typical hybrid nanotechnology applications.

The ANM-oxDNA/oxRNA hybrid models are intended to help design and probe function of large nucleic-acid protein hybrid nanostructures, but also to be used to study biological complexes and processes which can be captured within the approximations employed by the models. As an example of the model's use, we show simulations of DNA-protein hybrid nanocage, DNA wrapped around histone, and a nascent RNA strand inside polymerase.
\begin{figure}[ht]
\centering
\includegraphics[width = 3.5in]{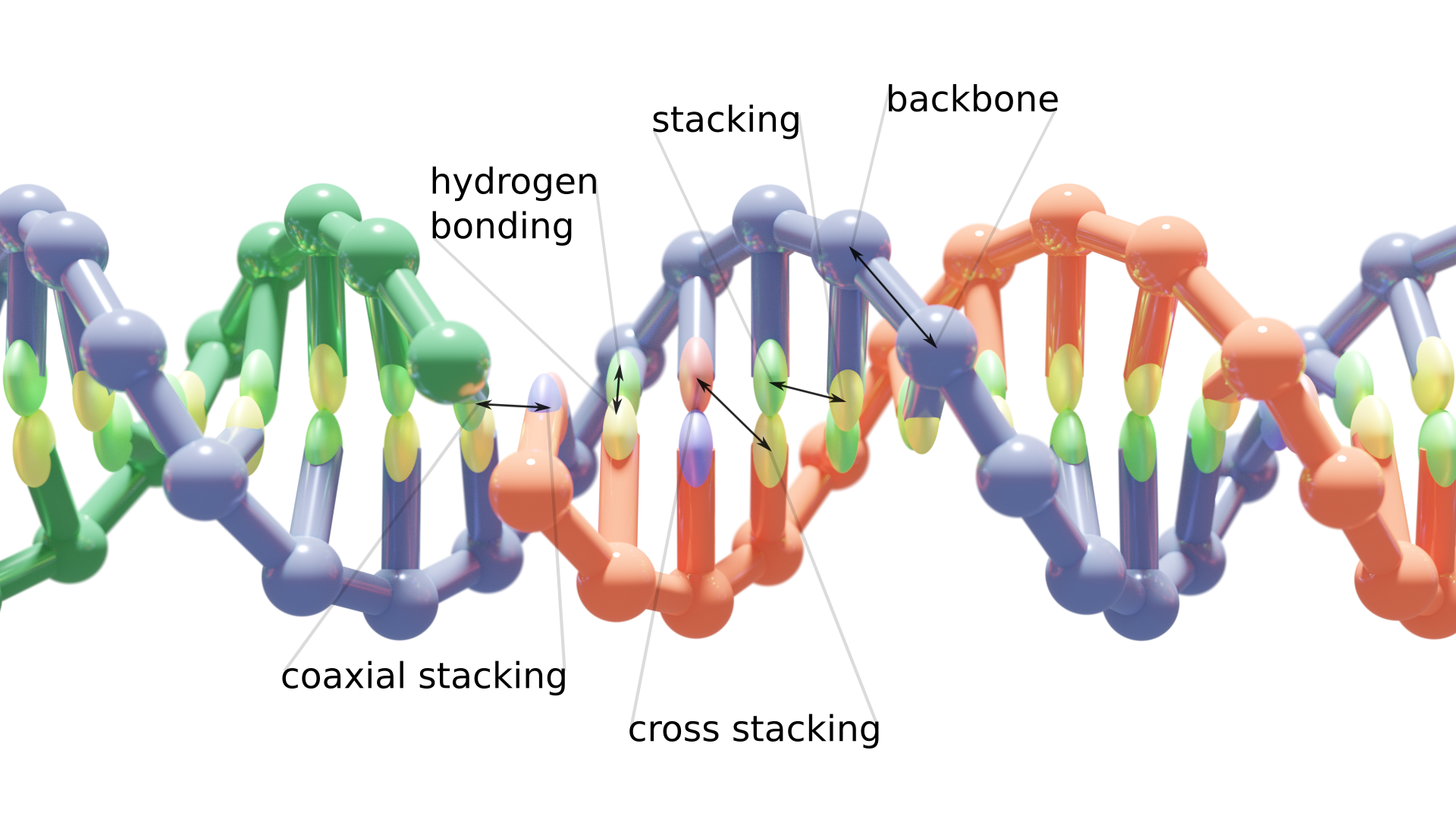}
\caption{A schematic overview of the oxDNA2 model and its interactions. Each nucleotide is represented as a single rigid body with backbone and base interaction sites (shown here schematically as a sphere and an ellipsoid) with their effective interactions designed to reproduce basic properties of DNA.}
\label{oxdna}
\end{figure}
\section*{Model Description}
Implemented in the oxDNA simulation package \cite{rovigatti2015comparison}, our model allows for a coarse-grained simulation of large hybrid nanostructures. It consists of two coarse-grained particle representations, the already existing oxDNA2 or oxRNA model for their respective nucleic acids and an Anistropic Network Model (ANM) for proteins \cite{atilgan2001anisotropy}. The detailed description of the oxDNA2/oxRNA models is available in Refs.~ \cite{snodin2015introducing,vsulc2014nucleotide}. A DNA duplex with a nicked strand is schematically illustrated in Fig.~\ref{oxdna}. The ANM allows us to represent a protein with a known structure as beads connected by springs. We chose to use ANM to represent proteins for its efficiency and relative simplicity, while still providing reasonably accurate representations of proteins crosslinked to DNA nanostructures. Furthermore, it can be implemented using only pairwise interaction potentials, the same as oxDNA/oxRNA models.

\subsection*{Protein Model}
\begin{figure}[ht]
\centering
\includegraphics[width = 3.5in]{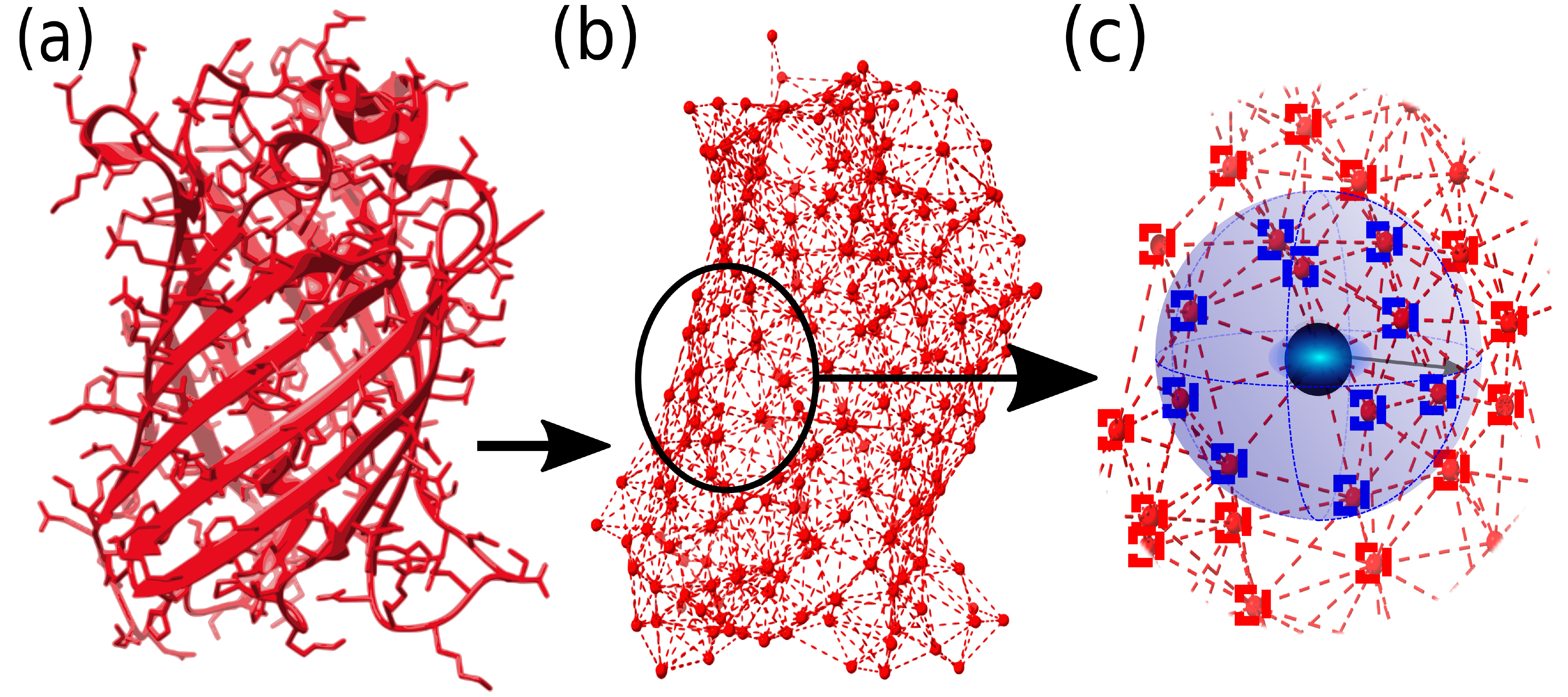}
\caption{Illustration of ANM using GFP protein (PDB code: 1W7S) from (a) starting PDB structure to (b) ANM representation at $r_{\rm max}$ of $\SI{8}{\angstrom}$, (c) bonding criteria per residue: all particles within distance $r_{\rm max}$ (bounds depicted by blue sphere) of center particle (black circle) are considered 'bonded' (blue squares) while those further (outside of sphere) are considered 'nonbonded' (red squares).}
\label{fig:anmcartoon}
\end{figure}

In the ANM representation, each protein residue is represented solely by its $\alpha$-carbon position. All residues within a specified cutoff distance $r_{\rm max}$ from one another are considered 'bonded'. Please see Ref.~ \cite{Atilgan2001} for a more detailed introduction. Each bond between residues $i$ and $j$ in the ANM is represented as a harmonic potential that fluctuates around the equilibrium length $r^{ij}_0$:
\begin{equation}
V_{ij} \left(r^{ij} \right) = \frac{1}{2} \gamma \left(r^{ij} - r^{ij}_{0} \right)^{2}
\label{v_ij}
\end{equation}
\noindent
 The total bonded interaction potential $V_{\rm bonded-anm}$ is the sum of terms  Eq.~\eqref{v_ij} for all pairs $i,j$ of aminoacids at distance smaller than $r_{\rm max}$ in the resolved protein structure, as schematically illustrated in Fig.~\ref{fig:anmcartoon}. We set $r^{ij}_0$ to the the distance between $\alpha$-carbons of the residues $i$ and $j$ in the PDB file.
Free parameter $\gamma$ is set uniformly on each bond in the ANM and and is chosen to best fit the Debye-Waller factors of the original PDB structure. Debye-Waller factors (or B-factors when applied specifically to proteins) describe the thermal motions of each resolved atom in a protein given by their respective X-ray scattering assay. As previously done \cite{Atilgan2001}, we use the B-factor of the $\alpha$-carbon to approximately capture the fluctuations of the protein backbone.   
Since an ANM is typically an analytical technique, it has no excluded volume effects. Hence we here extend the model to use a repulsive part Lennard-Jones potential between both bonded and non-bonded particles (Eq.~\ref{eq_repulsion}) to model the excluded volume at a per particle excluded volume diameter of $\SI{2.5}{\angstrom}$ .\\
%
%
\begin{table}[]
\small
  \caption{Excluded volume parameters used in Eq.~\ref{eq_repulsion} for (a) protein-protein, (b) protein-nucleic base and (c) protein-nucleic backbone non-bonded interactions in simulation units.}
  \label{tbl:xvparams}
  \begin{tabular*}{0.48\textwidth}{@{\extracolsep{\fill}}llll}
    \hline
    Parameter & (a) & (b) & (c)\\
    \hline
    $\sigma$ & $0.350$ & $0.360$ & $0.570$\\
    $r_c$ & $0.353$ & $0.363$ & $0.573$ \\
    $r^*$ & $0.349$ & $0.359$ & $0.569$\\
    $b$ & $30.7 \times 10^7$ & $29.6 \times 10^7 $& $17.9 \times 10^7$\\
    \hline
  \end{tabular*}
\end{table}
%
%
For any two particles (either protein/protein or protein-DNA/RNA) that are at distance $r$, we define the excluded volume interaction in Eq.~\ref{eq_repulsion}:
\begin{equation}
    V_{\rm exc}(r) = 
    \begin{cases}
        4  \epsilon  (-\frac{\sigma^{6}}{r^6} + \frac{\sigma^{12}}{r^{12}}) & r< r^* \\
        b  \epsilon  (r - r_c)^4 & r^* < r < r_c \\
        0 & r\geq r_c \\
    \end{cases}
    \label{eq_repulsion}
\end{equation}
\noindent
where we choose $r_c$.
Parameters $b$ and $r^*$ were calculated so that $V_{\rm exc}$ is a differentiable function. The constant $\epsilon$ sets the strength of the potential and we use $\epsilon = 82 {\rm \frac{pN}{nm}}$.
%
%
\subsubsection*{Parameterization}
In parameterizing our model for simulation, the goal is to mimic the dynamics of the protein in the native state. Though not without their setbacks \cite{Sun2019,Fuglebakk2013} we selected B-factors for their widespread availability in PDB structures and history of being used to fit elastic network models of proteins \cite{Sun2019}. Our model contains two free parameters, the cutoff distance  $r_{\rm max}$ and the spring constant $\gamma$. 

\begin{figure}[]
\centering
\includegraphics[width = 3.5in]{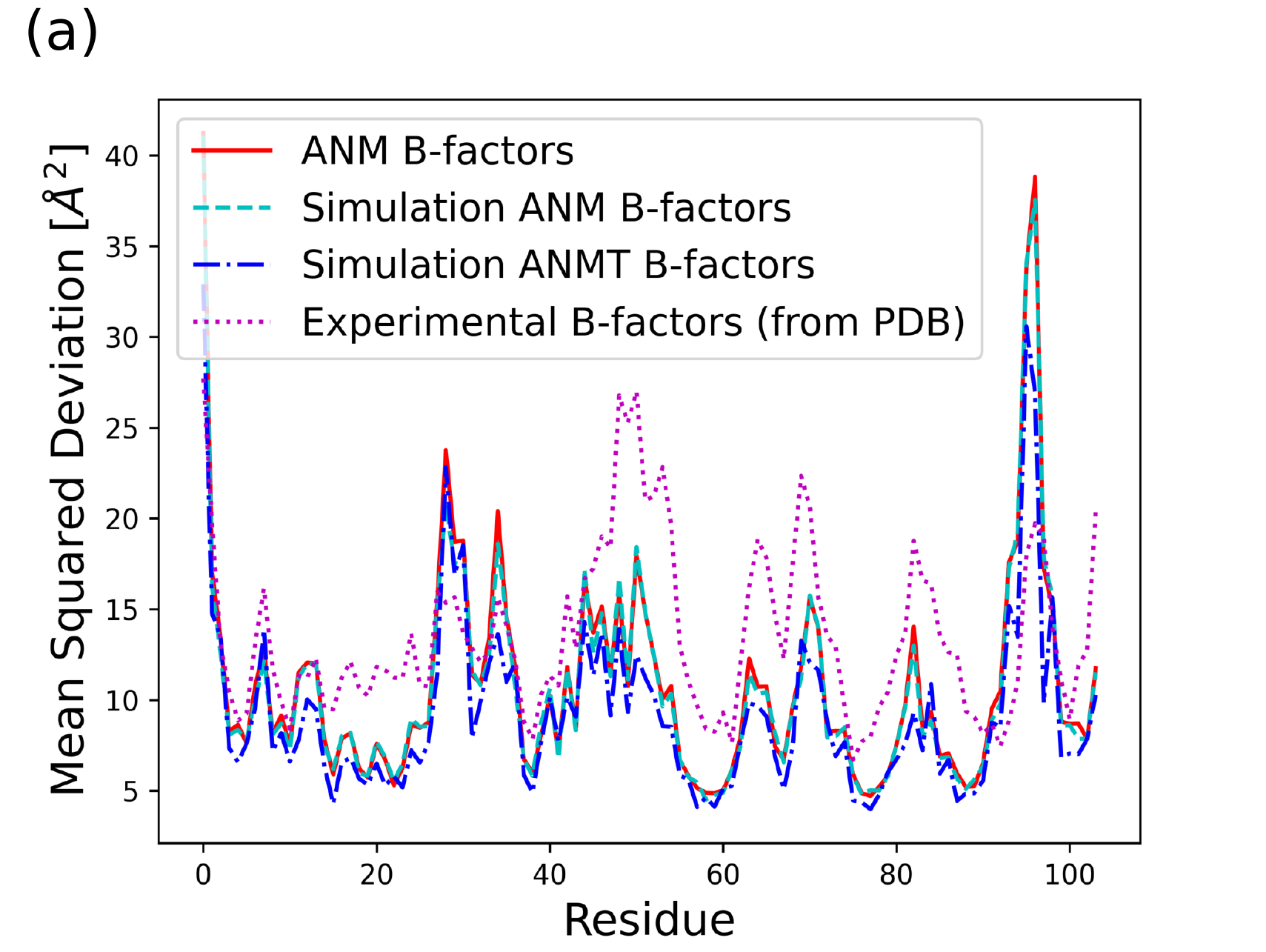}
\includegraphics[width=3.5in]{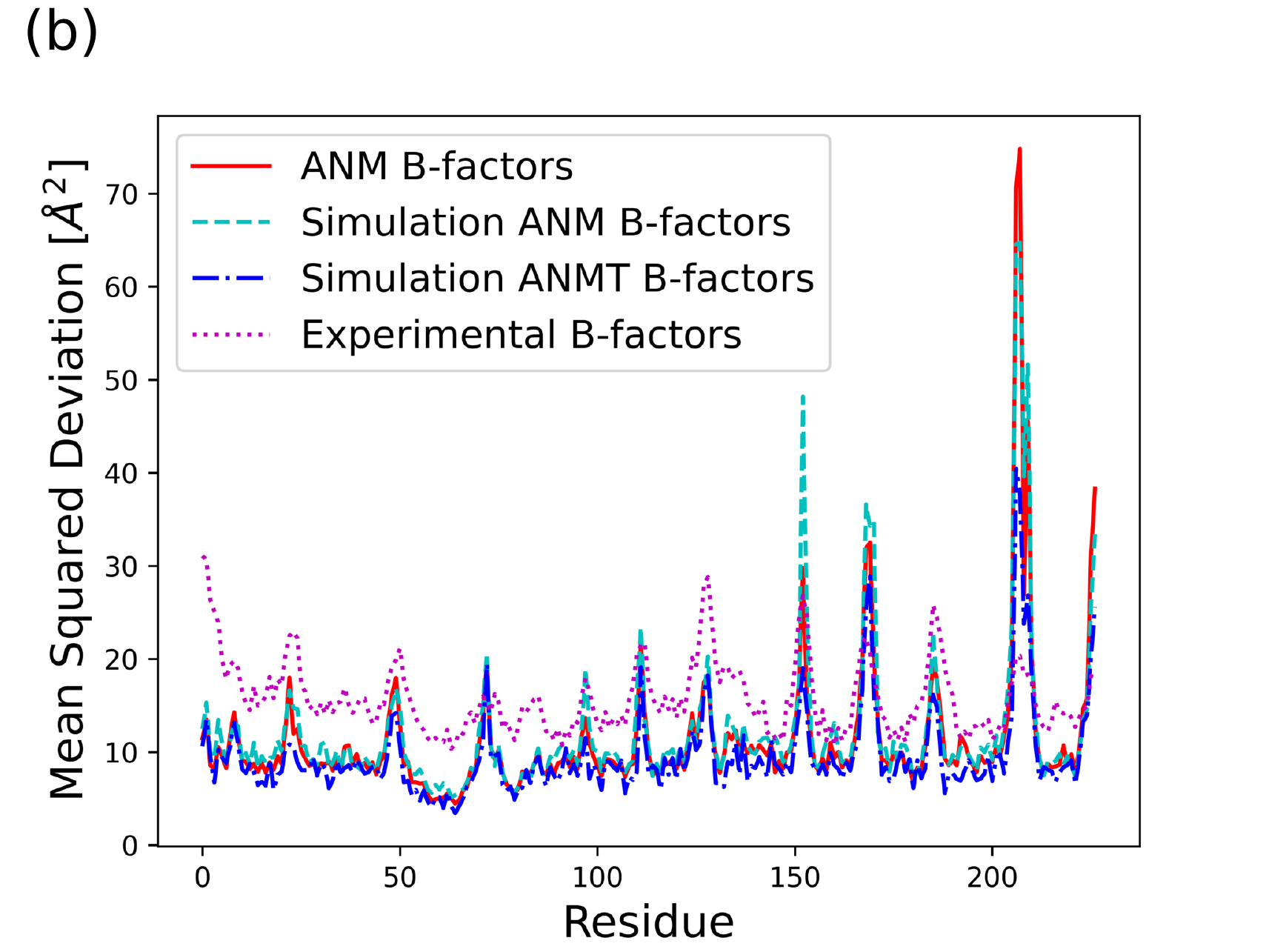}
\caption{Analytical, classic ANM simulation, ANMT simulation, and experimentally determined B-factors calculated in $\SI{}{\angstrom}^{2}$ per residue for \emph{(a)} ribonuclease T1 (PDB code 1BU4) at 25$\degree C$ ($r_{max} = 15, k_s = 42.2 pN/\SI{}{\angstrom}, k_b = k_t = 171.3 pN/\SI{}{\angstrom}$) and \emph{(b)} green fluorescent protein (PDB code 1W7S) at 25$\degree C$ ($r_{max} = 13, k_s = 33.2 pN/\SI{}{\angstrom}, k_b = k_t = 171.3 pN/\SI{}{\angstrom}$)}
\label{fig:proteinbfactor}
\end{figure}

The cutoff distance $r_{\rm max}$ can be varied,  but typically authors use a value of $\SI{13}{\angstrom}$ \cite{Atilgan2001} with strong long-range interactions being a key feature of the classic ANM \cite{Fuglebakk2013}. 
For each protein (consisting of $N$ aminoacids) represented by ANM, we linearly fit the analytically computed B-factors to their experimental counterpart with $\gamma$ as a free parameter. To solve for the B-factors analytically, we first calculate the $3N \times 3N$ Hessian matrix of the spring potential $V_{\rm spring}$, a task made simple by the harmonic potential energy function \cite{Atilgan2001}. After constructing the Hessian $H$ for the system at a specified cutoff $r_{\rm max}$, the mean squared deviation from the mean position for each residue $i$ can be calculated from the equipartition theorem:
\begin{equation}
\left\langle \Delta R_i^2 \right\rangle = \frac{k_b T}{\gamma} \left( tr\left( H^{-1}_{i,i} \right) \right)
\label{eq_hess}
\end{equation}
The B-factor $B$ of the residue $i$ can be directly computed from our previous result as \cite{Atilgan2001}:
\begin{equation}
B_i = \frac{8 \pi^2}{3} \left\langle \Delta R_i \right\rangle^2 .
\label{b_factor}
\end{equation}
The experimental B-factors are provided along with resolved crystal structures of proteins, and we can hence use Eqs.~\eqref{eq_hess} and \eqref{b_factor} to obtain $N$ equations. We then fit $\gamma$ parameter to minimize 
\begin{equation}
f(\gamma) =    \sum_{i=1}^N \left( B^{\rm exp.}_i - \frac{8\pi^2}{3} \left\langle \Delta R_i \right\rangle^2   \right)^2
\end{equation}
for a selected $r_{\rm max}$.
%
We can further measure the mean square deviation of residue positions in a simulation of our model and compare to the analytical calculation. We show the comparison in Fig.~\ref{fig:proteinbfactor} for ribonuclease T1 and green fluoresecent proteins simulated with the ANM model and our ANMT model, to be introduced later. While the simulation and analytical prediction of the classic ANM agree well with each other, as expected, we note that the model still does not fully reproduce the measured B-factors as reported in the experimental structures. ANM models are not able to fully reproduce the measured B-factors \cite{Atilgan2001}, and are known to have peaks in the mean square displacement profiles that have not been observed in the measured B-factors \cite{Sun2019}. The model nevertheless provides semi-quantitative agreement with the measured data, and hence represents an accurate enough representation of a protein to model its mechanical properties under small perturbations, as required for DNA-hybrid nanotechnology systems.
%
%
%

\section*{Expansion of the ANM model}
In addition to the classic ANM model, our model can also optionally use unique $\gamma_{ij}$ for each bonded pair of residues, which allows for implementation of other analytical models, such as the heterogeneous ANM (HANM) \cite{Xia2013} and multiscale ANM (mANM) \cite{Xia2017} that can generate better fits to experimental B-factors using the $\gamma_{ij}$ values. The HANM iteratively fits a normal ANM network to given experimental B-factors with variable realistic force parameters $\gamma_{ij}$. While unquestionably useful, the inaccuracy of B-factor data particularly in large or high resolution structures limits its application. In the mANM model, our conversion from the PDB structure to ANM representation also allows the fitting of multiple networks with varying $\gamma_{ij}$ values tuned by scale parameters \cite{Xia2017} (similar to $r_{\rm max}$). A linear combination of the networks is then solved to minimize the difference between the ANM network's predicted and experimental B-factors. The original formulation of the mANM \cite{Xia2017} is limited in computational application as it has no cutoff value ($r_{max}$); a protein of size $N$ residues would have $N(N-1)/2$ connections, significantly more than the average ANM. For the  proteins studied in this work, neither HANM nor mANM provided a significant advantage, so we decide to use the simple ANM with fixed $r_{\rm max}$ and the same $\gamma$ for all spring interactions. A C$\alpha$ coarse-grained HANM and a mANM with an additional cutoff value parameter are however implemented in our conversion scripts and can be optionally used to represent proteins in our model.

One major obstacle in using an ANM is known as the tip effect \cite{Lu2006}. The result is an extremely large spike in the B-factors due to a residue being under-constrained. Often this can be solved by raising the cutoff value in ANM construction; however, doing so raises the computational requirements of our simulations.  Furthermore, we found the ANM model to be not able to accurately represent short peptides, as the spring network does not provide enough constraints to reproduce their end-to-end distance as seen when simulated with more detailed models like AWSEM-MD \cite{Tsai2016}.

%
%

%
To overcome this obstacle, we implemented harmonic pairwise bending and torsional modulation forces into the existing simulation model. These new constraints allow for reduced $r_{max}$ values, and also can more accurately represent shorter peptides, which are often used in DNA-hybrid nanostructures. We introduce these optional modulation forces below. 
\subsection*{Bending and Torsional Modulation}
\begin{figure}
    \centering
    \includegraphics[width=3.5in]{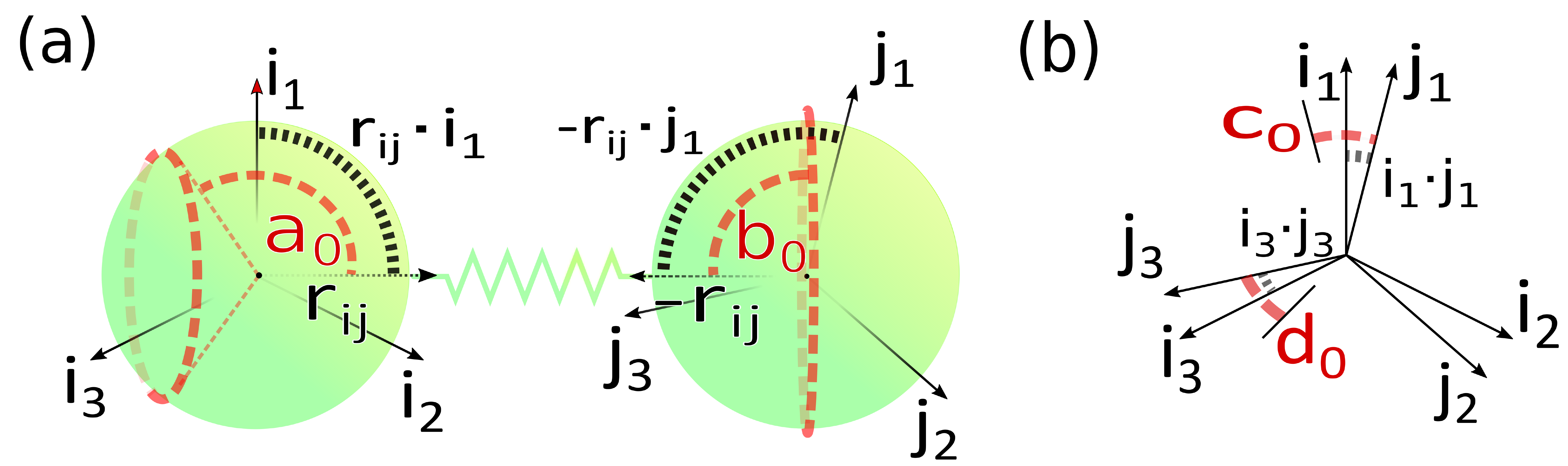}
    \caption{Depiction of (a) bending and (b) torsional potential terms on a pair of particles $i$ and $j$. The angles depicted as dot products correspond to the cosine of that angle. Equilibrium values (in red) correspond to (the cosine of) initial angle displacements derived from coordinates in the PDB file.}
    \label{fig:kbtsketch}
\end{figure}

We introduce the torsional and bending potential as optional interaction potentials in our protein representation on top of the ANM model with bonded and excluded volume 
potentials. Each protein residue corresponds to a spherical particle, with associated orientation given by its orthonormal axes $\hat{i}_1$, $\hat{i}_2$, $\hat{i}_3$ (Fig.~\ref{fig:kbtsketch}a). 
 Harmonic terms control the angle between the normalized  interparticle distance vector $\hat{r}_{ij}$ and the normal vector of each particle $\hat{i}_1, \hat{j}_1$ to control bond bending. The angles between two sets of orientation vectors, ${\hat{i}_1, \hat{j}_1}$ and ${\hat{i}_3, \hat{j}_3}$, are controlled as well allowing for modulation of the torsion based on the particles relative orientations. The full pairwise potential is given by Eq.~\ref{eqn:anmtpotential}:
\begin{multline}
    V_{ij}^{B\&T} = \frac{k_b}{2} \left( \left(\hat{\textbf{r}}_{ij} \cdot \hat{\textbf{i}}_1 - a^{ij}_0 \right)^2 + \left(-\hat{\textbf{r}}_{ij} \cdot \hat{\textbf{j}}_1 - b^{ij}_0\right)^2 \right) +\\
     \frac{k_t}{2} \left( \left(\hat{\textbf{i}}_1 \cdot \hat{\textbf{j}}_1 - c^{ij}_0 \right)^2 + \left(\hat{\textbf{i}}_3 \cdot \hat{\textbf{j}}_3 - d^{ij}_0 \right)^2 \right)
    \label{eqn:anmtpotential}
\end{multline}
\noindent
The function $V_{ij}^{B\&T}$ is defined for all pairs of residues that are neighbors along the protein backbone.
We set the energy minimum values $a^{ij}_0, b^{ij}_0, c^{ij}_0, d^{ij}_0$ to correspond to the cosines of respective angles in between residues in the PDB file for the protein structure. The terms $k_b$ and $k_t$ are two new global parameters that control the strength of the bending and torsion potential respectively. Currently, we set their values empirically, though pair specific terms could lead to further agreement with experimental data. Fig.~\ref{fig:proteinbfactor} shows the effect of the torsional and bonding modulation on the same set of proteins used prior. As intended, a noticeable decrease in high peak B-factors is observed using a modest $k_b$ and $k_t$ value. Fig.~\ref{fig:kbtsketch} illustrates the potential in a two particle system. Hereafter, we will refer to the ANM model with torsional and bending modulation as the ANMT model. 

%
\subsection*{Protein-Nucleic Acid Interactions}
In our current implementation of the model, protein residues and nucleotides have no interaction except for excluded volume and optional explicitly specified spring potentials between user-designated protein residues and nucleotides:
\begin{equation}
    V_{\rm spring}(r) = k \left(r - r_0 \right)^2
    \label{eq:spr}
\end{equation}
where $r$ is the distance between the centers of mass of the respective particles and $k$ and $r_0$ and external parameters.

The excluded volume interaction potential between protein and DNA/RNA residues has the same form as defined in Eq.~\eqref{eq_repulsion}, with the respective interaction parameters given in Table \ref{tbl:xvparams}. In the oxDNA/oxRNA models, each nucleotide has two distinct interaction sites (backbone and base), each of which is interacting with the protein residue using separate excluded volume parameters.
Future expansion of the model will include an approximate treatment of electrostatic interaction between protein and nucleic acids based on Debye-H\"{u}ckel theory as implemented in oxDNA \cite{snodin2015introducing}, as well as coarse-grained protein model AWSEM \cite{Tsai2016}. 
Many non-specific DNA-protein interactions make use of the electrostatic interactions between the DNA backbone and positively charged portions of the protein \cite{Misra1998}. Sensitive to salt concentration, these electrostatic contributions have been previously modeled using Debye-H\"{u}ckel theory \cite{Marcovitz2013} to investigate the role of protein frustration in regulating DNA binding kinetics. Similarly an extension of our model with an appropriate Debye-H\"{u}ckel potential can capture and enable study of non-specific DNA-binding protein systems.

\begin{figure}
    \centering
    \includegraphics[width=3.4in]{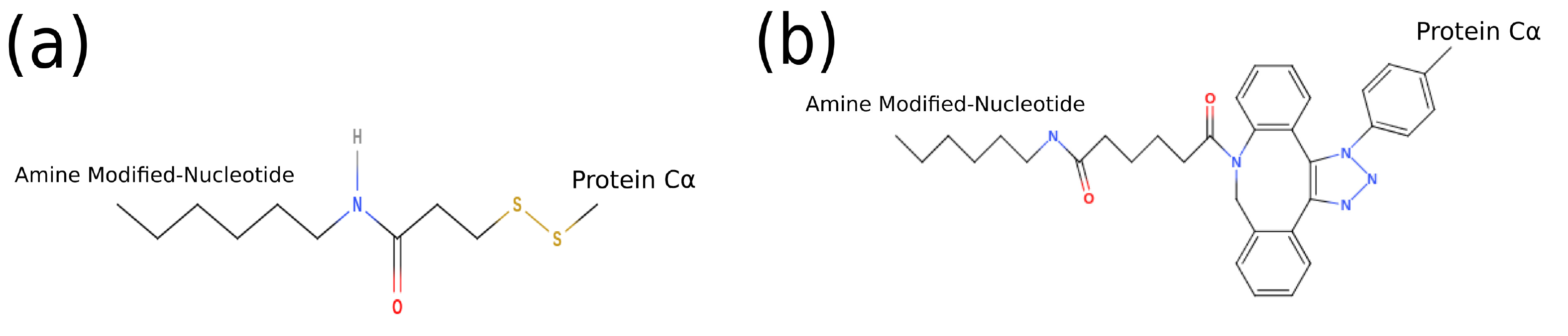}
    \caption{2D molecular structures of common bioconjugate linkers dubbed (a) LC-SPDP and (b) DBCO-triazole; both can be used to conjugate proteins to amine-modified nucleotides}
    \label{fig:linkers}
\end{figure}

Since we are interested in exploring conjugated hybrid systems, it is necessary to have an approximation for the covalent linkers bridging the nucleic acid base and protein residue. We model the two bioconjugate linkers, LC-SPDP and DBCO-triazole, (Fig.~\ref{fig:linkers}) that are typically used in protein-DNA hybrid nanotechnology \cite{buchberger2019hierarchical,xu2019tunable} using a spring potential as defined in Eq.~\eqref{eq:spr} with parameters $k$ and $r_0$ parametrized to mimic the end-to-end average distance and standard deviation of each linker at temperature 300K. LC-SPDP links the thiol group of a modified cysteine residue to an amine-modified nucleotide. DBCO-trizaole is the product of a copper-free click reaction involving a DBCO-modified residue to link to an azide-modified nucleotide. Each of the linkers (Fig.~\ref{fig:linkers}) was first drawn in MolView and then converted into OPLS/AA forcefield format via LibParGen \cite{Dodda2017,Dodda2017a,Jorgensen2005}. In GROMACS \cite{Berendsen1995}, each linker was first equilibrated and then simulated with OPLS/AA forcefield in SPCE water molecules at 300K for 3 trials of 1 nanosecond each. The obtained averaged end-to-end distance and standard deviation during each trial are shown in Table~\ref{tbl:linkers}. 
\begin{table}[h]
\small
  \caption{\ Average and standard deviation of end-to-end distance of linkers in fully atomistic Gromacs simulation and fit spring constant $k$}
  \label{tbl:linkers}
  \begin{tabular*}{0.48\textwidth}{@{\extracolsep{\fill}}lllll}
    \hline
    Linker & $\langle r \rangle$ ($\SI{}{\angstrom}$) & $\langle r^2 \rangle$ $(\SI{}{\angstrom})$ & $k$ ($pN / \SI{}{\angstrom}$)\\
    \hline
    LC-SPDP & 1.71 & 0.32 & 3.99\\
    DBCO-triazole & 3.64 & 2.87 & 0.14\\
    \hline
  \end{tabular*}
\end{table}
%
%
\section*{Examples}
Our model is fully functional with the latest version of the visualization tool oxView \cite{Poppleton2020} for both the design of hybrid nanomaterials as well as the viewing of simulation trajectories. The one caveat is that protein topologies are non-editable. Instead each protein starts from their PDB crystal structure and is converted into oxDNA format while the ANM spring constant is set to best match the experimental B-factors via our provided scripts. The output files can then be loaded into oxView as well as used for simulation in our model.

The model is theoretically able to represent any protein or protein complex that the ANM model can represent. Not beyond the scope of our model, biologically relevant multi-chain proteins such as nucleosomes, RNA polymerases, and viral assemblies can be also simulated, allowing for the nucleic acid behavior present in each of these systems to be modeled, studied, and compared to experimental data. While the detailed study of these systems is beyond the scope of this article, we show examples of both biological systems and designed nanosystems as represented by our ANM-oxDNA or ANM-oxRNA model.

\begin{figure*} 
 \centering
 \includegraphics[height=6cm]{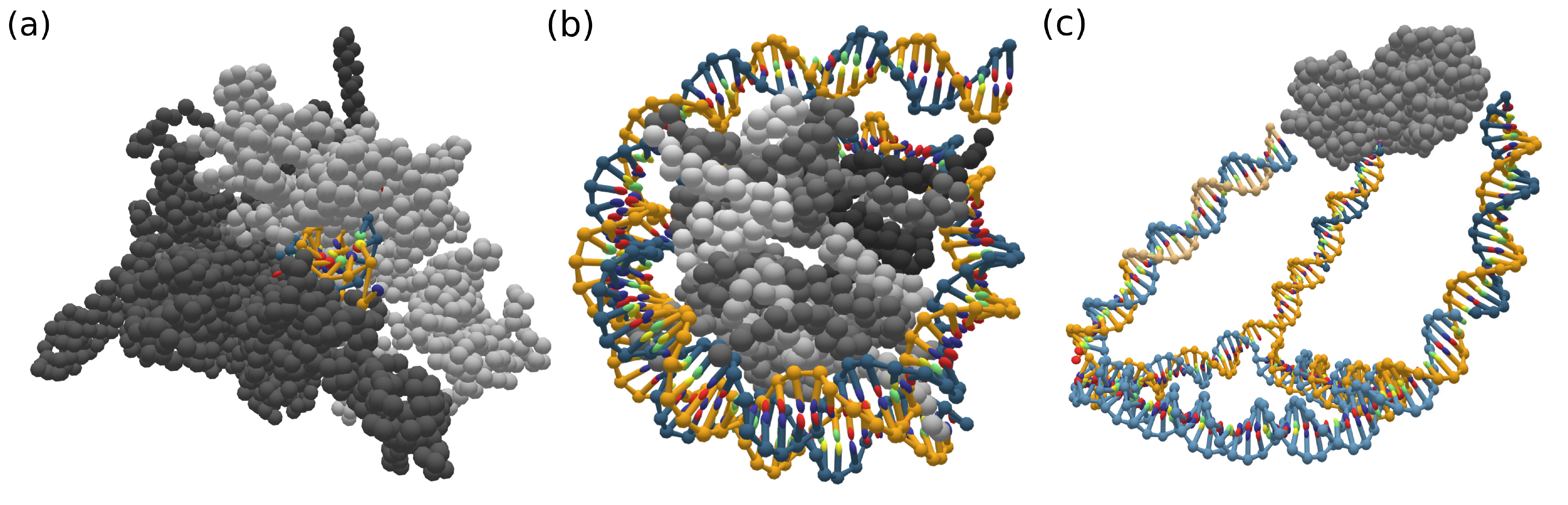}
 \caption{OxView visualization of simulated biological assemblies (a) RNA in exit channel of paused RNA polymerase (PDB code: 6ASX) and (b) Human nucleosome made up of histone octamer and DNA (PDB code: 3LEL), (c) mean structure from MD simulation of KDPG aldolase (PDB code: 1WA3) conjugated to a DNA cage}
 \label{fig:multichain}
\end{figure*}

Two prominent cases of nucleic acid - protein interactions, RNA polymerases and nucleosomes, were constructed and simulated using the ANMT model for future study. As many PDB files are missing residues, we first reconstruct each individual chain using the best scoring of ten models generated by the Modeller tool \cite{Sali1993}. The reconstructed RNA polymerase was converted into oxDNA format from its PDB entry (6ASX) using an $r_{max}$ of $\SI{15}{\angstrom}$. A fragment of the RNA was reinserted into the exit channel and the subsequent MD simulation was allowed to sample the RNA's escape from the exit channel. The reconstructed nucleosome was converted into oxDNA simulation format from its PDB entry (3LEL) using an $r_{max}$ of $\SI{12}{\angstrom}$. Attractive spring potentials between randomly chosen DNA and protein residues that were in close proximity in the PDB structure were added at the histone/DNA interface. 
Example snapshots from the MD simulations of these simulated biological systems are shown in Fig.~\ref{fig:multichain}a,b. 

While no process was explicitly modeled, our new model can be used to explore behavior of large scale simulation of DNA and histones, as at the latest version of GPU cards, the oxDNA model has been shown to be able to equilibrate systems consisting of over 1 million nucleotides. 
More pertinent to our goal of aiding in the design of hybrid nanostructures, our model supports conversion of CadNano, Tiamat, and other popular DNA origami design tools into the oxDNA format \cite{taco} where they can easily be edited in oxView to include linked proteins of interest. Since an ANM is a highly simplified model of protein dynamics, the predictive power of our model lies not in prediction of protein structure but rather the collection of statistical data of the protein's effect on the nucleic acid component of the system. Available and compatible with this model is also the suite of oxDNA analysis scripts \cite{Poppleton2020} allowing for a detailed exploration of system specific effects. 

Synthetic peptides are used in many chemistry applications. Since these peptides are often very small and lack long-distance contacts that enforce specific 3D conformations, we wanted to explore how our models perform on these small structures. We compared the end-to-end distance of 3 hemagglutinin binding peptides \cite{johnston2017simple} simulated in our ANM model, the ANMT model, and another popular coarse-grained protein model, AWSEM-MD \cite{davtyan2012awsem}.  For AWSEM-MD simulations, initial structure predictions were generated from sequence using I-TASSER \cite{yang2015tasser}. A secondary structure weight (ssweight) file was generated using jpred \cite{drozdetskiy2015jpred4}, and the structure and weight files were converted to the appropriate formats for AWSEM-MD simulation in LAMMPS \cite{plimpton1995fast} using tools provided with AWSEM-MD.  Simulations were run for $10^9$ steps with end-to-end distance printed every $10^5$ steps. 

Using the classic ANM, each peptide was built using strong backbone connections and significantly weaker long-range connections to empirically match the AWSEM mean and standard deviation of the end-to-end distance. The resulting simulation of each peptide; however, showed the trajectory to include a large amount of stretched, nonphysical conformations. The subsequent inclusion of the bending and torsion modulation using the ANMT model allowed for the same level of accuracy using only strong short-range connections. The ANMT model showed much higher rigidity with no stretched conformations when compared to the ANM model alone.  Final end-to-end distances and standard deviation are shown in Table \ref{tbl:pepdata}.

\begin{table}[h]
\small
  \caption{\ Average and standard deviation of end-to-end distance of hemagglutinin peptides between coarse-grained models}
  \label{tbl:pepdata}
  \begin{tabular*}{0.48\textwidth}{@{\extracolsep{\fill}}llll}
    \hline
    Model & AWSEM & ANM & ANMT\\
    \hline
    \emph{Peptide 125} & & &\\
    $\langle r \rangle$ ($\SI{}{\angstrom}$) & 12.02 & 12.9 & 12.09 \\
    $\langle r^2 \rangle$ ($\SI{}{\angstrom}$) & 4.9 & 4.51 & 4.34 \\
    \hline
    \emph{Peptide 149} & & &\\
    $\langle r \rangle$ ($\SI{}{\angstrom}$) & 12.9 & 12.9 & 12.9 \\
    $\langle r^2 \rangle$ ($\SI{}{\angstrom}$) & 6.6 & 4.6 & 4.6 \\
    \hline
    \emph{Peptide 227} & & &\\
    $\langle r \rangle$ ($\SI{}{\angstrom}$) & 14.5 & 16.2 & 14.7 \\
    $\langle r^2 \rangle$ ($\SI{}{\angstrom}$) & 7.4 & 5.4 & 5.1 \\
    \hline
  \end{tabular*}
  \begin{flushleft}
  \footnotesize{\emph{Peptide 125 - CSGHNIYAQYGYPYDHMYEG}}\\
  \footnotesize{\emph{Peptide 149 - CSGKSQEIGDPDDIWNQMKW}}\\
  \footnotesize{\emph{Peptide 227 - CSGSGNQEYFPYPMIDYLKK}}
  \end{flushleft}
\end{table}

Hybrid DNA-protein nanostructure constructs such as those developed by the Stepahanopoulos Lab are of particular interest. The Stephanopoulos group has experimentally realized their size-tunable DNA cage attached to homotrimeric protein KDPG aldolase making use of a LC-SPDP linker (Fig.~\ref{fig:linkers}) to join the DNA and protein components \cite{Xu2019}. The DNA cage was converted from Tiamat format into oxDNA format and the protein was converted from it's PDB structure. The linker between the components was modeled as a spring potential (Eq.~\eqref{eq:spr}) using the parameters from Table \ref{tbl:linkers}. We conducted a short MD simulation of the full system corresponding to time of about~30 ns. The mean structure from simulation of the experimental cage was calculated using our analysis scripts \cite{Poppleton2020} and is displayed in Fig.~\ref{fig:multichain}c.
\section*{Conclusions}
We present a coarse-grained protein model, based on elastic network representation of proteins, for use in conjunction with existing coarse-grained nucleic acid models capable of simulating large hybrid nanostructures. Implemented on GPU as well as CPU, our model allows for simulations of large systems 
based on nanotechnology designs as well as large biological complexes.

Looking forward, both the paused RNA polymerase and histone are biological systems we plan to study using this model. In addition, experimental systems such as the hybrid cage in Fig.~\ref{fig:multichain} can be simulated and directly compared to available experimental data. While widely available, B-factors are severely limited particularly in terms of accuracy. However, our model can be parameterized to approximate any available fluctuation data including but not limited to fully atomistic simulation and solution NMR data. In addition to the model, we also extended a nanotechnology design and simulation analysis tool, oxView, to include a protein representation to aid computer design of DNA/RNA-protein hybrid nanostructures. The subsequent analysis of the designs can be used to optimize nanostructure parameters, such as placement of the linkers and lengths of duplex segments in order to achieve desired geometry.

The simulation code is freely available on \url{github.com/sulcgroup/anm-oxdna} and will also be incorporated in the future release of the oxDNA simulation package. The visualization of protein-hybrid systems has been incorporated into our previously developed oxView tool \cite{Poppleton2020}. The aforementioned analysis scripts and visualizer are available in git repositories  \url{github.com/sulcgroup/oxdna_analysis_tools} and \url{github.com/sulcgroup/oxdna-viewer} respectively. 
%

\section*{Acknowledgements}
 We thank all members of the \v{S}ulc group for their support and helpful discussions, in particular to H.~Liu and M.~Matthies. We thank Dr.~Stephanopoulos for helpful comments and feedback about simulation study of his DNA-protein hybrid system. 
We acknowledge support from the NSF grant no.~1931487.




\bibliography{soft_matter.bib} 

\end{document}